\documentclass[aps,prl,reprint,amsmath,amssymb,showpacs,superscriptaddress]{revtex4-1}
\usepackage{CJK}
\usepackage{graphicx}
\usepackage{dcolumn}
\usepackage{bm}
\usepackage{color}
\usepackage{hyperref}

\begin{document}

\title{Next-to-leading-order prediction for the neutrinoless double-beta decay}
\author{Y. L. Yang}
\affiliation{State Key Laboratory of Nuclear Physics and Technology, School of Physics, Peking University, Beijing 100871, China}

\author{P. W. Zhao}
\email{pwzhao@pku.edu.cn}
\affiliation{State Key Laboratory of Nuclear Physics and Technology, School of Physics, Peking University, Beijing 100871, China}

\begin{abstract}
The neutrinoless double-beta decay ($0\nu\beta\beta$) of two neutrons $nn\rightarrow ppee$ is the elementary subprocess of $0\nu\beta\beta$ decay in nuclei.
Accurate knowledge of the  $nn\rightarrow ppee$ amplitude is required to pin down the short-range contributions in the nuclear matrix elements of the candidate nuclei for large-scale $0\nu\beta\beta$ searches.
In this Letter, we report the first next-to-leading-order prediction of the $nn\rightarrow ppee$ amplitude, with Bayesian uncertainty quantification.
This is made possible by the development of the relativistic chiral effective field theory, in which no unknown contact term is required up to next-to-leading order.
The theory is validated by reproducing in a parameter-free way the available data on the charge independence and charge symmetry breaking contributions in the two-nucleon scattering.
The present work makes an essential step towards addressing the uncertainty in the theoretical calculations of the nuclear matrix elements relevant for $0\nu\beta\beta$ searches.
\end{abstract}

\maketitle

{\it Introduction.}---The search for neutrinoless double-beta decay ($0\nu\beta\beta$) is one of the top priorities in the field of nuclear and particle physics~\cite{Aalseth2018Phys.Rev.Lett.132502,Adams2020Phys.Rev.Lett.122501,Agostini2020Phys.Rev.Lett.252502,
Albert2018Phys.Rev.Lett.072701,Armengaud2021Phys.Rev.Lett.181802,Arnold2017Phys.Rev.Lett.041801,Gando2016Phys.Rev.Lett.082503,Dai2022Phys.Rev.D32012}.
Its observation would signal the violation of the lepton number conservation of the Standard Model, prove the Majorana nature of neutrinos~\cite{Schechter1982Phys.Rev.D29512954,Zeldovich1981JETPLett.141}, constrain the neutrino mass scale and hierarchy~\cite{Avignone2008Rev.Mod.Phys.481}, and shed light on the matter-antimatter asymmetry in the Universe~\cite{Davidson2008Phys.Rep.105}.
However, the interpretation of current experimental limits and, even more so of the potential future discoveries, is hampered by substantial uncertainties in the theoretical calculations of the nuclear matrix elements.
Within the standard picture of $0\nu\beta\beta$ involving the long-range Majorana neutrino exchange~\cite{Weinberg1979Phys.Rev.Lett.15661570}, as we consider here, various calculations of nuclear matrix elements lead to discrepancies larger than a factor of three~\cite{Engel2017Rep.Prog.Phys.046301,Yao2022Prog.Part.Nucl.Phys.126.103965,Agostini2023Rev.Mod.Phys.025002}.

The uncertainties of the nuclear matrix elements stem from two aspects: first, from the approximated nuclear many-body wave functions of the initial and final nuclear states, and second, from the uncertainty of the $0\nu\beta\beta$ decay operator.
The latter has recently received wide attention from the community, as nuclear effective field theory (EFT) analyses~\cite{Cirigliano2018Phys.Rev.Lett.202001,Cirigliano2019Phys.Rev.C055504} have revealed a potential leading-order (LO) contribution from a previously unrecognized contact decay operator.
Various nuclear-structure calculations indicate that its contribution could be comparable to the standard long-range nuclear matrix elements~\cite{Cirigliano2018Phys.Rev.Lett.202001,Wirth2021Phys.Rev.Lett.242502,Jokiniemi2021Phys.Lett.B136720,Belley2024Phys.Rev.Lett.182502}.
However, the size of the contact coupling is highly uncertain due to the absence of lepton-number-violating data.
This introduces a significant uncertainty in the calculations of the nuclear matrix elements and, thus, in the interpretation of the large-scale searches for $0\nu\beta\beta$ decay.

The $nn\rightarrow ppee$ process, which is the elementary subprocess of $0\nu\beta\beta$ in nuclei, plays a key role in addressing the uncertainty of the contact decay operator.
A first-principle calculation of the $nn\rightarrow ppee$ process from lattice QCD is being pursued energetically by the community~\cite{Cirigliano2020Prog.Part.Nucl.Phys.103771,Drischler2021Prog.Part.Nucl.Phys.121.103888,Davoudi2021Phys.Rep.174}.
However, this is extremely challenging, although significant progress has been made in calculating the pionic amplitude of the $0\nu\beta\beta$ decay~\cite{Feng2019Phys.Rev.Lett.022001,Tuo2019Phys.Rev.D094511,Detmold2020arXiv2004.07404,Nicholson2018Phys.Rev.Lett.121.172501} and the double-beta decays at unphysical pion masses~\cite{Shanahan2017Phys.Rev.Lett.062003,Tiburzi2017Phys.Rev.D054505,Davoudi2024Phys.Rev.D114514}.
Alternatively, the $nn\rightarrow ppee$ amplitude was modeled with the generalized Cotthingham formula~\cite{Cirigliano2021Phys.Rev.Lett.172002,Cirigliano2021J.HighEnergyPhys.289}.
This approach introduces phenomenological inputs for elastic intermediate states and neglects inelastic contributions, resulting in model-dependent uncertainties that need to be further scrutinized.
In contrast, a relativistic chiral EFT has been developed to predict the $nn\rightarrow ppee$ amplitude~\cite{Yang2024Phys.Lett.B855.138782}. 
It avoids phenomenological inputs such as off-shell $NN$ amplitudes, and its uncertainty can be estimated by the truncation errors of the chiral expansion.

In this Letter, we present the first next-to-leading-order (NLO) prediction of the $nn\rightarrow ppee$ amplitude using the relativistic chiral EFT, with uncertainty quantified by the Bayesian approach.
This is possible because no unknown contact term is needed to renormalize the $nn\rightarrow ppee$ amplitude up to NLO in the present relativistic framework.
The theory is validated by reproducing in a parameter-free way the available data on the charge independence breaking (CIB) and charge symmetry breaking (CSB) of the nucleon-nucleon ($NN$) scattering.
The present results provide by far the most accurate estimate of the $nn\rightarrow ppee$ amplitude in the EFT framework and could thus greatly reduce the uncertainties in the nuclear matrix elements.

{\it Relativistic chiral EFT for $0\nu\beta\beta$ at NLO.}---
We focus on the $nn\rightarrow ppee$ process in the $^1S_0$ channel, which dominates at low energies and is the only channel relevant to the LO contact term.
Within chiral EFT, the $nn\rightarrow ppee$ amplitude is expanded in powers of the small expansion parameter $Q/\Lambda_\chi$ with the soft scale set by the pion mass $Q\sim m_\pi$ and the chiral symmetry breakdown scale set by the lightest vector meson $\Lambda_\chi\sim m_\rho$~\cite{Weinberg1990Phys.Lett.B288},
\begin{equation}\label{eq.exp}
  \mathcal{A}_{fi}=\mathcal{A}_{fi}^{(0)}+\mathcal{A}_{fi}^{(1)}+\mathcal{A}_{fi}^{(2)}+\ldots.
\end{equation}
Here, the superscript denotes the power of $Q/\Lambda_\chi$ [we label LO as O(1)],  and the subscripts $f$ and $i$ denote the final $pp$ and initial $nn$ states, respectively.
The NLO contribution $\mathcal{A}_{fi}^{(1)}$ to the amplitude is induced by the effective range correction in the $^1S_0$ channel~\cite{Long2012Phys.Rev.C024001,Cirigliano2019Phys.Rev.C055504}.

In the present framework, the two-nucleon processes are described by the relativistic scattering equation
\begin{equation}\label{eq.BS}
  \begin{split}
    T(\bm p_f,\bm p_i)&=V(\bm p_f,\bm p_i)+\int\frac{{\rm d}^3 k}{(2\pi)^3}V(\bm p_f,\bm k)G_0(\bm k;E)T(\bm k,\bm p_i),\\
    G_0(\bm k;E)&=\frac{M^2}{\bm k^2+M^2}\frac{1}{E+2M-2\sqrt{\bm k^2+M^2}+{\rm i}0^+},
  \end{split}
\end{equation}
where $E$ is the total kinetic energy, $M$ the nucleon mass, and $\bm p_f$ ($\bm p_i$) the two nucleons' final (initial) momentum in the center-of-mass frame.
This relativistic scattering equation, along with many others of a similar type that satisfy relativistic elastic unitarity, can be derived from a Lorentz-invariant three-dimensional reduction of the Bethe-Salpeter equation~\cite{Kadyshevsky1968Nucl.Phys.B125, Woloshyn1973Nucl.Phys.B269}. However, in the present framework, it is obtained by employing time-ordered perturbation theory without nonrelativistic reduction to calculate the $NN$ scattering amplitudes~\cite{Epelbaum2012Phys.Lett.B338344}.
For the $^1S_0$ channel, the strong potential up to NLO takes the form
\begin{equation}\label{eq.Vs}
    \begin{split}
        V_S(\bm p',\bm p)=C-\frac{g_A^2}{4f_\pi^2}\frac{m_\pi^2}{m_\pi^2+\bm q^2}
        +\left(\delta C+D\frac{\bm p'{}^2+\bm p^2}{2}\right)
    \end{split}
\end{equation}
with the momentum transfer $\bm q=\bm p'-\bm p$, the axial coupling $g_A=1.27$, the pion decay constant $f_\pi=92.2$ MeV, and the pion mass $m_\pi=138.03$ MeV.
Here, the LO part includes the momentum-independent contact term $C$ and the pion-exchange Yukawa potential, and the NLO part includes a shift $\delta C$ to the LO contact term and a new momentum-dependent contact term $D$ responsible for the effective range correction.
To achieve renormalization beyond LO, the NLO corrections are included perturbatively on top of the LO contributions~\cite{vanKolck1999Nucl.Phys.A273,Beane1998Nucl.Phys.A445,Long2012Phys.Rev.C024001}, where the LO potential is iterated nonperturbatively in Eq.~(\ref{eq.BS}).
There are no unknown low-energy constants (LECs) in the weak sector up to NLO, since the neutrino potential up to NLO is contributed only by the long-range neutrino exchange~\cite{Cirigliano2018Phys.Rev.C065501},
\begin{equation}\label{eq.Vnu}
  V_\nu(\bm q)=\frac{\tau_1^+\tau_2^+}{\boldsymbol q^2}\left\{1+2g_A^2\left[1+\frac{m_\pi^4}{2(\boldsymbol q^2+m_\pi^2)^2}\right]\right\}.
\end{equation}

\begin{figure}[!htpb]
    \centering
	\includegraphics[width=0.48\textwidth]{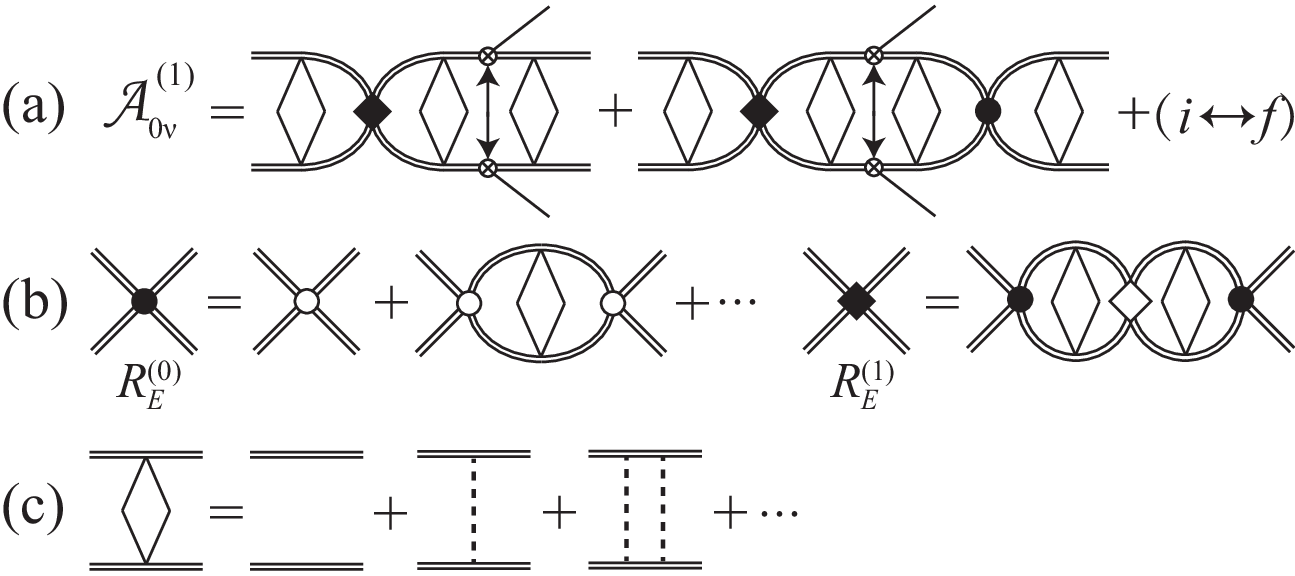}
	\caption{Diagrams representing (a) the full NLO correction to the $nn\rightarrow ppee$ amplitude, (b) the renormalization of the LO (circles) and NLO (squares) contact couplings, $R_E^{(0)}$ and $R_E^{(1)}$, (c) the Yukawa propagator generated by the resummation of the pion-exchange potential.
The double, plain, and dashed lines denote nucleon, lepton, and pion fields, respectively.
The double-headed arrow denotes an insertion of neutrino potential $V_\nu$.
The empty circle (square) denotes an insertion of the LO (NLO) bare contact coupling.
The parenthesis ($i\rightarrow f$) in panel (a) denotes the diagrams similar to the two shown but with the initial and final states exchanged.
}\label{fig1}
\end{figure}
The diagrams contributing to the NLO amplitude $\mathcal{A}_{fi}^{(1)}$ are shown in Fig.~\ref{fig1} (a).
They are expressed in terms of the renormalized couplings $R_E^{(0)}$ and $R_E^{(1)}$ [Fig.~\ref{fig1} (b)] and the Yukawa propagator $G_\pi$ resumming the pion-exchange potential [Fig.~\ref{fig1} (c)].
Specifically, the renormalized NLO $nn\rightarrow ppee$ amplitude reads~\cite{Supp}
\begin{equation}\label{eq.Afi}
  \begin{split}
     \mathcal{A}_{fi}^{(1)}&=\chi_{E_f} R^{(1)}_{E_f}J_{fi}^{1\text{-loop}}+\chi_{E_f} R^{(1)}_{E_f}J_{fi}^{2\text{-loop}}R^{(0)}_{E_i}\chi_{E_i}+(i\leftrightarrow f),
  \end{split}
\end{equation}
with $E_{f(i)}$ the final-(initial-) state kinetic energy and
\begin{equation}\label{eq.coup}
  R^{(0)}_{E}=\frac{1}{C_R^{-1}-I_{0,R}(E)},\quad R^{(1)}_{E}=\frac{D_R}{2C_R^2}\frac{ME}{[C_R^{-1}-I_{0,R}(E)]^2}.
\end{equation}
Here, $J_{fi}^{1,2\text{-loop}}$ are the loop integrals with a neutrino exchange in the two diagrams in Fig.~\ref{fig1} (a), $\chi$ is the Yukawa wave function related to the semicircles dressed by the pion-exchange potential in the final or initial states~\cite{Kaplan1996Nucl.Phys.B629,Long2012Phys.Rev.C024001}.
The renormalized LECs $C_R$ and $D_R$ result from the absorption of the ultraviolet divergences of the bubble integrals into the bare coupling constants $C$, $\delta C$ and $D$,
\begin{equation}\label{eq.renorm}
  \begin{split}
     C_R^{-1}-I_{0,R}(E)&=\lim_{\Lambda\rightarrow\infty}\left[C_\Lambda^{-1}-I_{0,\Lambda}(E)\right],\\
     \frac{D_R}{2C_R^2} ME&=\lim_{\Lambda\rightarrow\infty}\left[\delta C_\Lambda C_\Lambda^{-2}+D_\Lambda C_\Lambda^{-1} I_{2,\Lambda}(E)\right],
  \end{split}
\end{equation}
where $\Lambda$ is the cutoff introduced in a regularization scheme, $I_{0,R}(E)=I_{0,\Lambda}(E)-I_{0,\Lambda}(0)$, and $I_{n,\Lambda}(E)=\int^\Lambda\frac{{\rm d}^3k}{(2\pi)^3} G_\pi(k;E) k^n$ the bubble integrals dressed by the pion-exchange potential.
On the left-hand side of Eq. (\ref{eq.renorm}), the renormalized LECs $C_R$ and $D_R$ are both finite and can be related to the scattering length and effective range, respectively, by matching the LO and NLO on-shell $NN$ amplitudes to the effective range expansion.
See Supplemental Materials for details on the renormalization of the LECs~\cite{Supp}.

In the present relativistic framework, the loop integrals $J_{fi}^{1,2\text{-loop}}$ are ultraviolet finite.
Following the renormalization analysis in Ref.~\cite{Yang2024Phys.Lett.B855.138782}, one finds $J_{fi}^{1,2\text{-loop}}$ are convergent as $O(\Lambda^{-2})$ when $\Lambda\rightarrow\infty$.
Therefore, the NLO amplitudes are properly renormalized without introducing additional contact terms, and are solely determined by the two renormalized LECs $C_R$ and $D_R$.
This is in stark contrast to the nonrelativistic heavy baryon approach, where the two-loop integral $J_{fi}^{2\text{-loop}}$ is logarithmic divergent when $\Lambda\rightarrow\infty$~\cite{Cirigliano2019Phys.Rev.C055504}.
Such a difference originates from the $1/M$ expansion of the propagator $G_0$ [Eq.~(\ref{eq.BS})], which in the nonrelativistic approach makes the ultraviolet behavior of the loop integrals more singular.
Similar improvements in renormalizability have also been found in nucleon-nucleon scattering~\cite{Epelbaum2012Phys.Lett.B338344} and few-body systems~\cite{Epelbaum2017Eur.Phys.J.A98,Yang2022Phys.Lett.B137587}.

\begin{figure}[!htpb]
    \centering
	\includegraphics[width=0.4\textwidth]{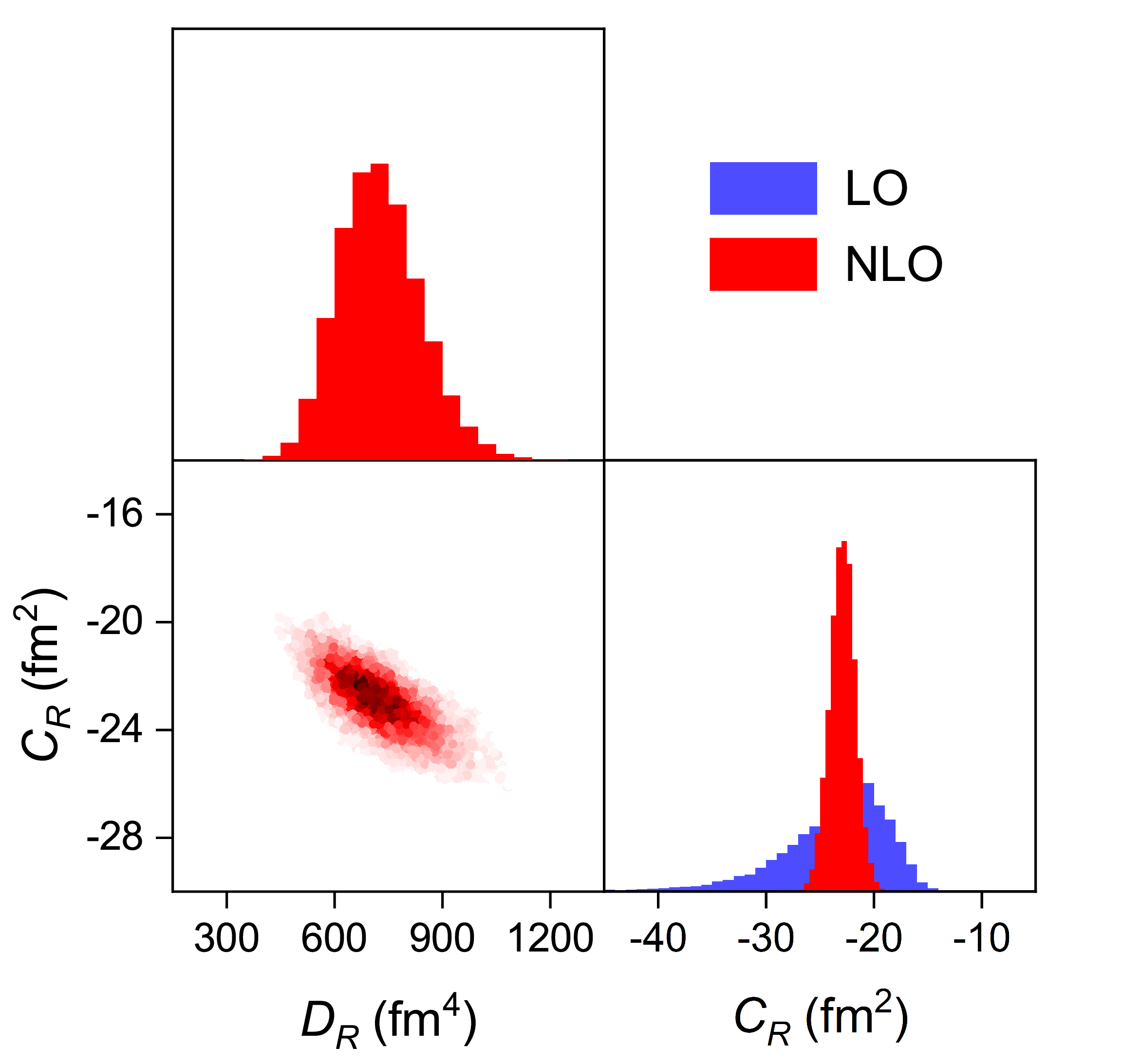}
	\caption{(Color online). Posterior probability density distributions for the renormalized low-energy constants $C_R$ and $D_R$.}\label{fig2}
\end{figure}

{\it Bayesian uncertainty quantification.}---Using the Bayes theorem, the posterior probability density function (PDF) of the renormalized LECs $\vec{\alpha}=(C_R,D_R)$ is determined by the likelihood ${\rm pr}(D|\vec{\alpha})$ and the prior ${\rm pr}(\vec{\alpha})$, 
\begin{equation}\label{eq.bayes}
  {\rm pr}(\vec{\alpha}|D)\propto {\rm pr}(D|\vec{\alpha}){\rm pr}(\vec{\alpha}).
\end{equation}
The LECs $C_R$ and $D_R$ are conditioned to the low-energy $np$ scattering data $D$, including the scattering length $a_{np}^{1S0}=-23.74(2)$ fm and the partial-wave cross sections $\sigma_{np}^{^1S_0}(E_{\rm lab})$ with $E_{\rm lab}\leq100$ MeV from the Nijmegen partial wave analysis~\cite{Stoks1993Phys.Rev.C48.792815}.
The prior is a Gaussian distribution that mildly encodes the expected scaling of the LECs.
The likelihood takes the form ${\rm pr}(D|\vec{\alpha})=\exp(-\chi^2/2)$, with the $\chi^2$ function estimated from the observables and their associated errors in the data set.
For a given observable, the error is a sum of the experimental error $\sigma^2_{\rm exp}$ and the theoretical error~\cite{Furnstahl2015Phys.Rev.C92.024005,Melendez2017Phys.Rev.C96.024003,Wesolowski2019J.Phys.G46.045102},
\begin{equation}\label{eq.sigmath}
  \sigma_{\rm th}^2=O_{\rm ref}^2\frac{(Q/\Lambda_\chi)^{2(\nu+1)}}{1-(Q/\Lambda_\chi)^2},
\end{equation}
where $O_{\rm ref}$ is a reference value, taken as the maximum likelihood estimation at LO, $Q=\max\{p,m_\pi\}$, $\Lambda_\chi=m_\rho=770$ MeV, and $\nu$ the chiral order. 
The posterior PDFs for the LECs are sampled using the Metropolis-Hastings algorithm~\cite{Metropolis1953J.Chem.Phys.21.1087,Hastings1970Biometrika57.97}, shown in Fig.~\ref{fig2}.
We adopt $10^4$ samples throughout this work.
One can see that the constraint on $C_R$ is significantly improved when moving from LO to NLO, and both $C_R$ and $D_R$ are well constrained at NLO.

{\it Validation.}---Before proceeding to the $0\nu\beta\beta$ amplitudes, we first validate our approach on the CIB and CSB of the $NN$ scattering data at low energies, which are controlled by the one-photon exchange process.
The motivation is that the one-photon exchange process has a similar structure to the $0\nu\beta\beta$ one, i.e., a massless propagator coupled to a two-nucleon matrix element of two hadronic currents.
The comparison with available $NN$ scattering data allows us to validate the predictions and the estimated uncertainties of the present approach.
To this end, we calculate the $nn$ and $pp$ scattering phase shifts using the potential in Eq.~(\ref{eq.Vs}) but with the neutral pion mass $m_{\pi^0}=134.98$ MeV and additional static Coulomb potential for the $pp$ case.
The LECs remain the same as those already determined in the $np$ scattering (Fig.~\ref{fig2}), since the charge-dependent contact terms appear only at higher orders~\cite{Epelbaum2009Rev.Mod.Phys.17731825}.
We have confirmed that the $nn$ and $pp$ phase shifts are indeed cutoff-independent as $\Lambda\rightarrow\infty$, as expected from the similar structure between neutrino and photon exchange.

\begin{figure}[!htpb]
    \centering
	\includegraphics[width=0.4\textwidth]{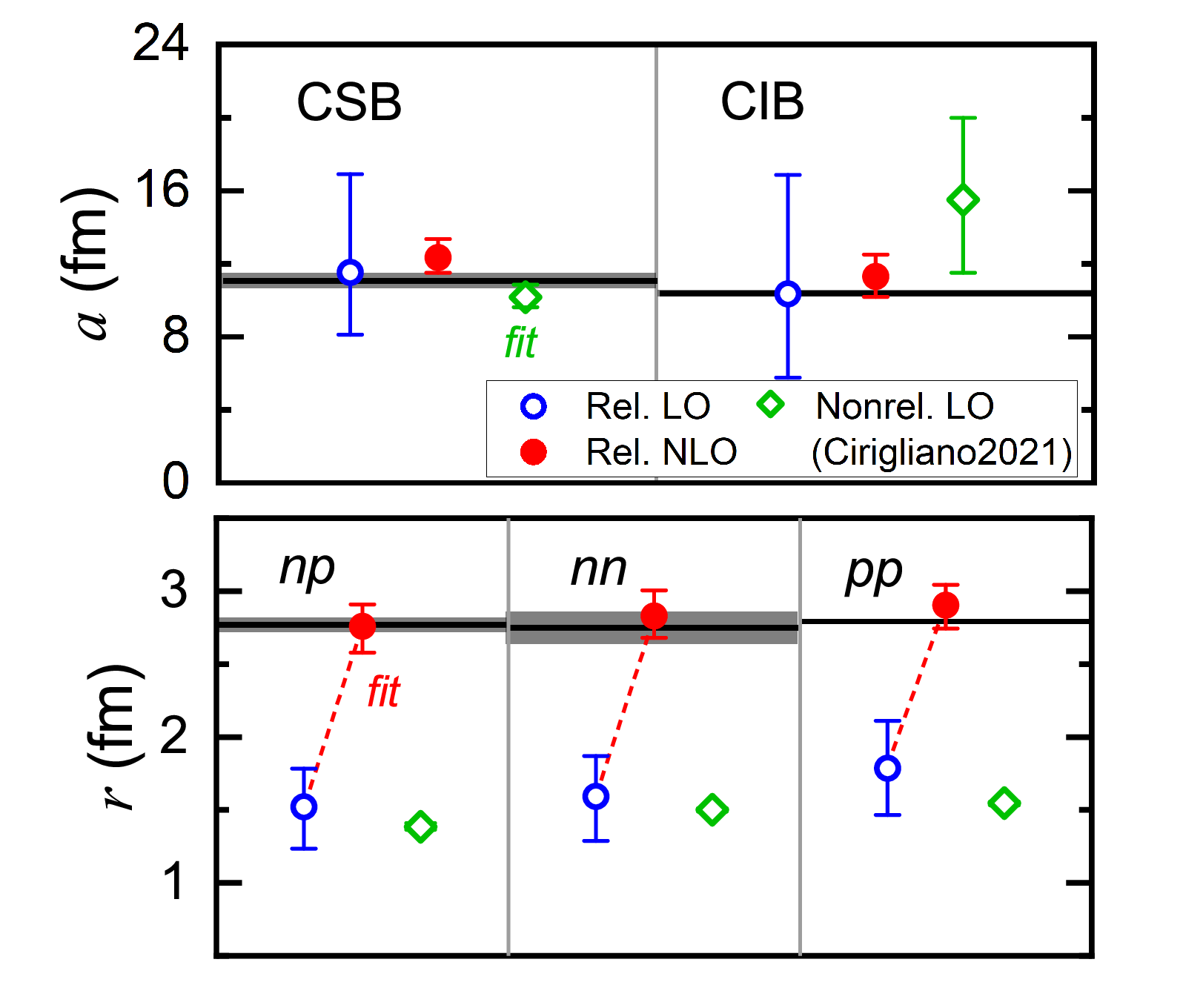}
	\caption{(Color online). The scattering lengths (upper panel) and effective ranges (lower panel) of $NN$ scattering predicted by the relativistic chiral EFT at LO and NLO.
The circles and error bars denote the maximum-likelihood estimates and the 68\% Bayesian confidence intervals, respectively.
Also shown are the nonrelativistic LO results constrained by the generalized Cottingham model by Cirigliano \textit{et al.}~\cite{Cirigliano2021Phys.Rev.Lett.172002, Cirigliano2021J.HighEnergyPhys.289}, and the errors are from the estimated uncertainties of the model assumptions (see text for details).
The black solid lines and bands denote the experimental data and errors~\cite{Stoks1993Phys.Rev.C48.792815,Chen2008Phys.Rev.C77.054002, Machleidt2001Phys.Rev.C63.024001,Miller1990Phys.Rep.194.1}.
}\label{fig3}
\end{figure}

Figure~\ref{fig3} depicts the scattering lengths $a$ and the effective range $r$ in the $^1S_0$ channel predicted by the relativistic chiral EFT at LO and NLO.
The scattering lengths are defined as
\begin{equation}\label{eq.a}
  a_{\rm CSB}=a_{pp}-a_{nn},\quad a_{\rm CIB}=\frac{a_{pp}+a_{nn}}{2}-a_{np}.
\end{equation}
The uncertainties include those from the LECs, sampled from their posterior PDFs, and those from the truncation in the chiral expansion, sampled from a normal distribution with width $\sigma_{\rm th}$ in Eq.~(\ref{eq.sigmath}).
The errors are given by the 68\% Bayesian confidence intervals~\cite{Bayesian}, corresponding to the 1$\sigma$ intervals for a normal distribution.

Already at LO, the relativistic chiral EFT predicts the CSB and CIB scattering lengths in agreement with the experimental values.
These predictions are further refined at NLO. The uncertainties are reduced from about 30\% to only 10\%, and the predicted $a_{\rm CSB}$ and $a_{\rm CIB}$ are still consistent with the experimental values.
Furthermore, the effective ranges for $np$, $nn$, and $pp$ scattering are all excellently described at NLO but are significantly underestimated at LO.
It is important to emphasize that there are no free parameters here, since the only two LECs are fixed by the $np$ scattering data.

In the nonrelativistic approach, the experimental $a_{\rm CIB}$ has also been used to test the generalized Cottingham model~\cite{Cirigliano2021Phys.Rev.Lett.172002,Cirigliano2021J.HighEnergyPhys.289}, and to estimate the size of the unknown LO contact term~\cite{Cirigliano2018Phys.Rev.Lett.202001,Jokiniemi2021Phys.Lett.B136720,Richardson2021Phys.Rev.C103.055501}.
In Fig.~\ref{fig3}, the nonrelativistic LO results including the CIB and CSB contact terms are also shown for comparison.
Here, the CIB contact term is determined by matching it to a vector amplitude provided by the generalized Cottingham model~\cite{Cirigliano2021Phys.Rev.Lett.172002, Cirigliano2021J.HighEnergyPhys.289}, while the CSB contact term is fitted to the experimental value of $a_{\rm CSB}$.
Therefore, only $a_{\rm CIB}$ is predicted from the generalized Cottingham model, with its lower boundary of the estimated uncertainty slightly above the experimental value.

\begin{figure}[!htpb]
    \centering
	\includegraphics[width=0.4\textwidth]{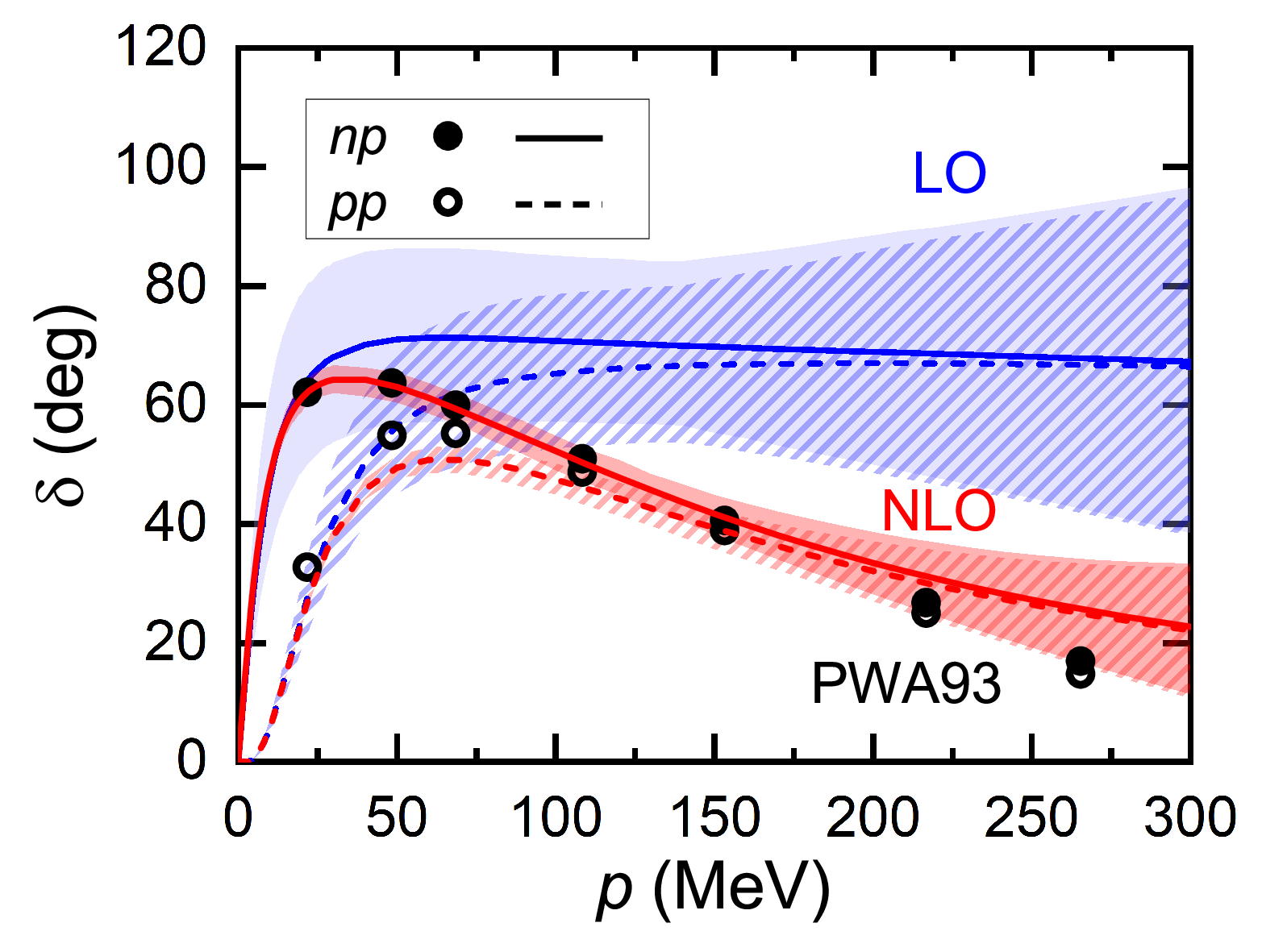}
	\caption{(Color online). The $np$ and $pp$ phase shifts predicted by the relativistic chiral EFT at LO and NLO, as functions of the center-of-mass momentum $p$.
    The lines and bands denote the maximum-likelihood estimates and the 68\% Bayesian confidence intervals, respectively.
    The solid and empty dots denote the data from the Nijmegen partial-wave analysis (PWA93)~\cite{Stoks1993Phys.Rev.C48.792815}.
}\label{fig4}
\end{figure}

To further illustrate the effective range physics introduced at NLO, Figure~\ref{fig4} depicts the $np$ and $pp$ scattering phase shifts predicted by the relativistic chiral EFT, compared to the data~\cite{Stoks1993Phys.Rev.C48.792815}.
At LO, both phase shifts are overestimated for large momenta, due to the missing of the effective range physics.
However, the situation improves significantly at NLO.
In particular, the splitting between the $np$ and $pp$ phase shifts is well described and the remaining small discrepancies should be attributed to higher-order effects.

The aforementioned analyses provide compelling evidence in support of the present relativistic chiral EFT approach and demonstrate that the corresponding Bayesian uncertainty estimates are realistic.

\begin{figure}[!htpb]
    \centering
	\includegraphics[width=0.45\textwidth]{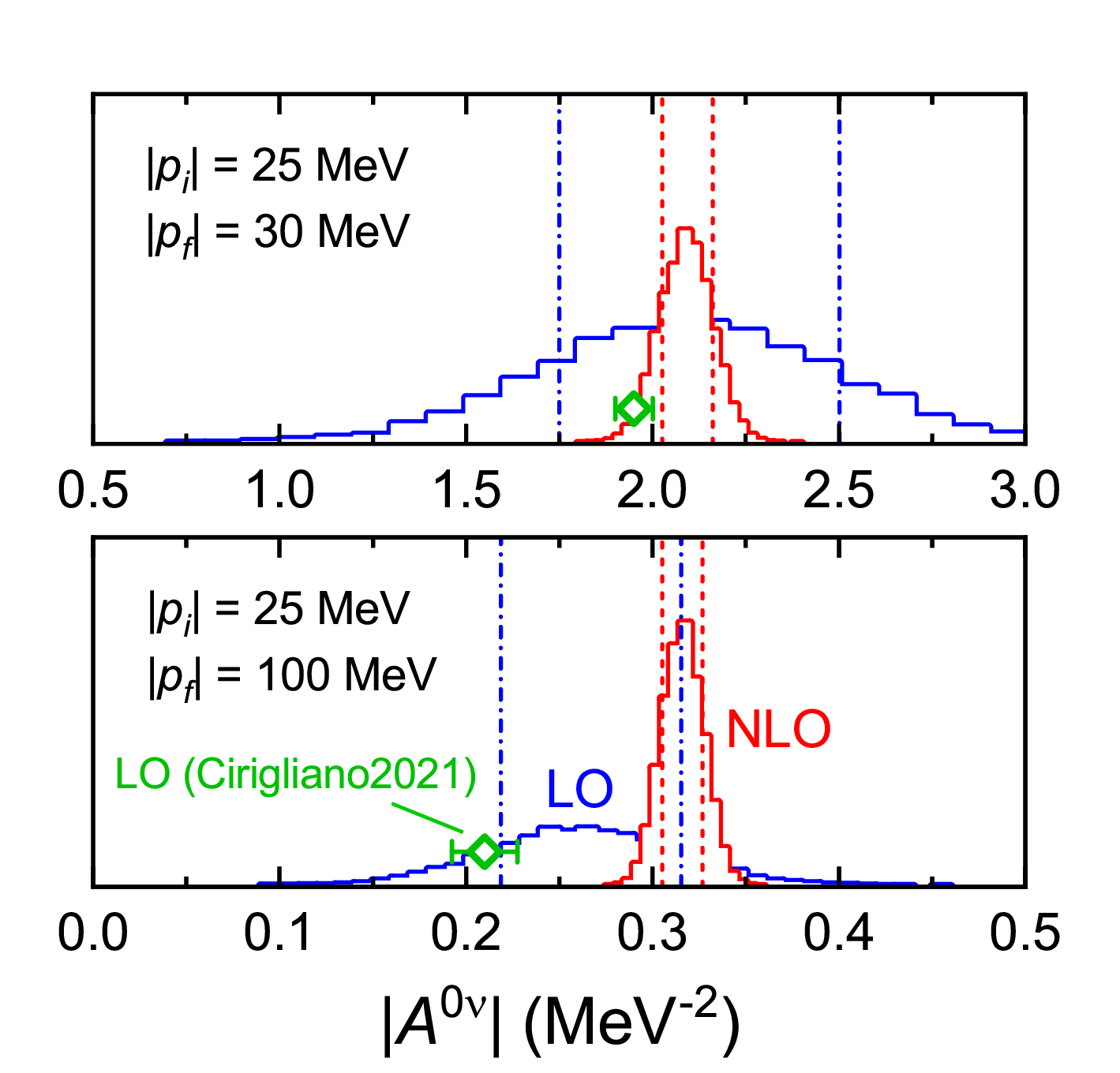}
	\caption{(Color online). Posterior probability density distributions for the $nn\rightarrow ppee$ amplitude at two different kinematics, obtained from the LO and NLO relativistic chiral EFT.
    The dot-dashed and dashed vertical lines indicate the 68\% Bayesian confidence intervals at LO and NLO, respectively.
The empty diamonds depict the results obtained using nonrelativistic LO chiral EFT with the contact term from the generalized Cottingham model by Cirigliano \textit{\textit{et al.}}~\cite{Cirigliano2021Phys.Rev.Lett.172002,Cirigliano2021J.HighEnergyPhys.289}.
Its error bar is an estimate of the uncertainties from the model inputs and assumptions.
}\label{fig5}
\end{figure}
{\it NLO predictions for $nn\rightarrow pp ee$ amplitudes.}---Figure~\ref{fig5} depicts the posterior PDFs for the $nn\rightarrow ppee$ amplitudes predicted by the relativistic chiral EFT at LO and NLO.
They are calculated from the sampled LECs and then added by truncation errors sampled from a normal distribution with width $\sigma_{\rm th}$ in Eq. (\ref{eq.sigmath}).
The present predictions of the $nn\rightarrow ppee$ amplitude are compared to the nonrelativistic LO results with the contact term from the generalized Cottingham model~\cite{Cirigliano2021Phys.Rev.Lett.172002,Cirigliano2021J.HighEnergyPhys.289}.
The LO nonrelativistic and relativistic results are consistent with each other within the estimated errors.
However, the NLO relativistic prediction is systematically larger than the nonrelativistic results.
At $|\bm p_f|=30$ MeV, the nonrelativistic result lies outside the 68\% Bayesian confidence interval of the NLO prediction, and is smaller than the maximum-likelihood estimate at NLO by about $10\%$.
At $|\bm p_f|=100$ MeV, the difference between the NLO relativistic and LO nonrelativistic results becomes more pronounced, about $50\%$.
It is understandable that the NLO correction becomes more significant at higher kinematics, since the chiral expansion is a low-momentum expansion.
In the future, it would be also interesting to investigate the higher-order inelastic corrections using the machinery developed recently in Ref.~\cite{VanGoffrier2025Phys.Rev.D055033}.

{\it Summary.}---We present the first NLO predictions for the $nn\rightarrow ppee$ amplitude, with Bayesian quantified uncertainties, using the relativistic chiral EFT approach.
We quote the NLO prediction
\begin{equation}\label{eq.amp}
  |\mathcal{A}_{0\nu}|=0.0209(7)\ {\rm MeV}^{-2}
\end{equation}
at the kinematics $|\bm p_i|=25$ MeV, $|\bm p_f|=30$ MeV.
The uncertainty comes from the estimation of the low-energy constants and the truncation in the chiral expansion.

The present result of the $0\nu\beta\beta$ amplitude has several merits compared to the previous model estimate at LO, $|\mathcal{A}_{0\nu}|=0.0195(5)$ MeV~\cite{Cirigliano2021Phys.Rev.Lett.172002, Cirigliano2021J.HighEnergyPhys.289}.
First, the present result is obtained within the chiral EFT framework and avoids phenomenological inputs such as off-shell $NN$ amplitudes.
Second, the present result includes, for the first time, Bayesian quantification of the uncertainty of the $nn\rightarrow pp ee$ amplitude arising from the low-energy constants and from neglecting subleading terms.
This allows a clear statistical meaning of the present result and is particularly in line with the recent theoretical endeavors of quantifying uncertainties in nuclear matrix elements of nuclei~\cite{Belley2024Phys.Rev.Lett.182502}.
Finally, the present relativistic EFT approach is stringently tested by reproducing, in a parameter-free way, the charge independence and charge symmetry breaking contributions in the nucleon-nucleon scattering amplitudes.
In Ref.~\cite{Yang2025Phys.Rev.D014507}, the relativistic EFT approach has also shown encouraging agreement with the recent first lattice QCD calculation~\cite{Davoudi2024Phys.Rev.D114514} of the $nn\rightarrow pp ee$ process, though at somewhat heavy pion mass.
Therefore, the present work provides the most accurate $0\nu\beta\beta$ amplitude to determine the contact decay operator in nuclear matrix elements.

\begin{acknowledgments}
We thank Evgeny Epelbaum and Yakun Wang for the valuable discussions.
This work has been supported in part by the National Natural Science Foundation of China (Grants No. 12141501, No. 123B2080, No. 12435006, No. 12475117, No. 11935003), National Key Laboratory of Neutron Science and Technology NST202401016, National Key R\&D Program of China 2024YFE0109803, and the High-performance Computing Platform of Peking University.
We acknowledge the funding support from the State Key Laboratory of Nuclear Physics and Technology, Peking University (Grant No. NPT2023ZX03).
\end{acknowledgments}

%

\end{document}